\documentclass[showpacs,twocolumn,prl]{revtex4}

\usepackage{graphicx}
\usepackage{rotating}
\usepackage{amsmath}
\usepackage{amsfonts}
\usepackage{amssymb}
\usepackage{graphicx, psfrag}
\usepackage{enumerate}
\usepackage{longtable}
\usepackage{dcolumn}% Align table columns on decimal point
\usepackage{bm}
\usepackage{epstopdf}
\usepackage[colorlinks=true, citecolor=blue, urlcolor = blue, linkcolor= red, bookmarks=true]{hyperref}

\begin{document}

\newcommand{\be}{\begin{equation}}
\newcommand{\ee}{\end{equation}}
\newcommand{\bn}{\begin{eqnarray}}
\newcommand{\en}{\end{eqnarray}}
\newcommand{\ii}{\'{\i}}
\newcommand{\ca}{\c c\~a}

\title{Quantum Critical Transport At A Continuous Metal-Insulator Transition}

\author{P. Haldar, M. S. Laad and S. R. Hassan}\email{prosenjit@imsc.res.in}

\affiliation{Institute of Mathematical Sciences, Taramani, Chennai 
600113 India}

\date{\rm\today}

\begin{abstract}
In contrast to the first-order correlation-driven Mott metal-insulator transition (MIT), 
continuous disorder-driven transitions are
intrinsically quantum  critical.  Here, we investigate transport quantum criticality in the Falicov-Kimball model, a representative of the latter class in the ``strong disorder'' category.  Employing cluster-dynamical mean-field theory (CDMFT), we find clear and anomalous quantum critical scaling behavior manifesting as perfect mirror symmetry of scaling curves on both sides of the MIT.  Surprisingly, we find that the beta-function,
$\beta(g)$, scales like log$(g)$ deep into the bad-metallic phase as well, providing sound unified basis for 
these findings.  We argue that such ``strong localization'' quantum criticality may manifest in real three-dimensional systems where disorder effects are more important than electron-electron interactions.
\end{abstract}
   

\maketitle

  The weak localization (WL) of non-interacting electrons due to disorder is now well understood within the scaling formalism~\cite{pwa1958pwa1979} as a genuine quantum phase transition.  In spite of its
extensive successes~\cite{LeeTVR1985}, further experimental developments~\cite{kravchenko,arindam} present compelling evidence for a different kind of quantum criticality that requires non-trivial
extensions of the WL paradigm.  It has long been suggested, both experimentally~\cite{rosenbaum} and more recently, theoretically~\cite{dobrosavljevic} that electron-electron interactions 
in a disordered system can cause a metal-insulator transition (MIT) in $D=2$ dimensions.  Another possibility is that the experiments may be probing the ``strong localization'' region of a disorder
model, $i.e$, in a regime $k_{F}l\leq 1$, opposite to that where WL theory works.  This is supported by the observation that features at odds with the WL predictions seem to be qualitatively similar
for $D=2,3$ systems~\cite{arindam}, as well as the fact that observed resistivities can greatly exceed $(2-3)\hbar/e^{2}$ (the Mott-Ioffe-Regel (MIR) limit), reaching unprecedentedly high values 
$O(500-700)\hbar/e^{2}$.  Further, beautiful ``mirror'' symmetry and associated scaling behaviors in transport, along with anomalous critical exponents suggestive of glassy freezing close to the MIT are known 
for the $2D$ electron gas in Si~\cite{kravchenko,arindam}.  In these cases, either of the two scenarios above can cause the perturbative approach underlying WL to break down.  This is because the infra-red pole
structure of the one-fermion propagator is supplanted by a branch-cut, putting the very notion of well-defined Landau-like quasiparticles in trouble in bad metals close to the MIT.  

   Such anomalous features as the above are also to be found in systems close to purely correlation-driven Mott transitions~\cite{terletska}.  For e.g, while resistivity curves ($\rho_{dc}(T,X)$, $X$ a control 
parameter, $e.g$, external pressure) weakly depend upon $X$ at high temperature $T$, they rapidly converge toward either metallic or insulating branches at low $T$.  The ``Mott'' quantum critical aspect is 
rather clearly borne out by beautiful scaling behavior and ``mirror'' symmetry of the scaling (beta) functions.  Since dynamical mean-field theory (DMFT) seems to capture this aspect for the Hubbard model, 
albeit only above the finite-$T$ critical end-point of the first-order MIT, the following issues arise: $(i)$ what kind of quantum criticality would operate if the MIT were to be continuous at $T=0$, and what 
scaling phenomenology should one then expect?  $(ii)$  What are its manifestations in transport in the quantum critical region?

  Though the possibility of ``strong localization'' has been studied~\cite{sudip1997} in the context of the MIT in $D=2$, no study of how such ``strong coupling'' quantum criticality might arise in transport in a 
specific microscopic model is yet available.  In this letter, we answer these questions for the spinless Falicov-Kimball model (FKM), which is isomorphic to the Anderson disorder model (ADM) with a binary 
alloy disorder distribution.  We choose the FKM since it can be exactly solved and shows a continuous MIT at $T=0$, both within DMFT and cluster-DMFT studies~\cite{freericks,ourfirstpaper,rowlands,jarrell}, 
allowing us to study the genuine quantum criticality in the ``strong localization'' limit in detail.  The Hamiltonian is~\cite{freericks}

\be
H_{FKM}= -t\sum_{<i,j>}(c_{i}^{\dag}c_{j} + h.c) + U\sum_{i}n_{i,c}n_{i,d}
\ee      
on a Bethe lattice with a semicircular band density of states (DOS) as an approximation to a $D=3$ lattice.  $c_{i}(c_{i}^{\dag}),d_{i}(d_{i}^{\dag})$ are fermion operators in dispersive band ($c$) 
and dispersionless ($d$) states, $t$ is the one-electron hopping integral and $U$ is the onsite repulsion for a site-local doubly occupied configuration.  Since $n_{i,d}=0,1$, $v_{i}=Un_{i,d}$ is also viewed as a static ``disorder'' potential for the $c$-fermions.  Since $H_{FKM}$ is exactly soluble in DMFT, extensive studies of dc and ac transport, relying on absence of vertex corrections in Bethe-Salpeter equations 
(BSE) for conductivities, have been done~\cite{freericks}.  The very interesting issue of the effects of inter-site correlations on transport in the FKM have, however, never been considered, to our best knowledge.  We use our recent exact-to-$O(1/D)$ extension of the DMFT for the FKM to investigate the issues $(i),(ii)$ detailed above, a program greatly facilitated by semi-analytic cluster propagators and 
self-energies~\cite{ourfirstpaper}.  This fortunate circumstance permits detailed analysis of transport properties near the ``Mott'' QCP in the FKM to $O(1/D)$ ($D$=spatial dimensionality), including situations with finite short-range order (SRO).  

   Remarkably, it turns out that transport properties can also be exactly computed in our 2-site CDMFT because: $(1)$ the ``bare'' bubble term in the BSE is directly obtained from the CDMFT Green functions
$G({\bf K},\omega)$ for cluster momenta ${\bf K}=(0,0).(\pi,\pi)$, its computation being most conveniently done in the real space bonding (S)-antibonding (P) cluster (C) basis, and $(2)$ more importantly, 
since the irreducible p-h vertex function $\Gamma({\bf K},\omega)$ identically vanishes, exactly as in DMFT.  This is because these enter in the form~\cite{kotliar} $\Gamma({\bf k})=\sum_{k}v_{k}*G_{k}*G_{k}$
 which identically vanish when ${\bf k}={\bf K}=(0,0),(\pi,\pi)$.  Thus, remarkably, only the bare bubbles contribute in both $S,P$ channels, and the total conductivity, 
$\sigma_{xx}(T)=\sigma_{xx}^{S}(T)+\sigma_{xx}^{P}(T)$, with 

\be
\sigma_{xx}(T)=\sigma_{0}\sum_{\bf K}\int_{-\infty}^{+\infty} d\epsilon v^2(\epsilon) \rho_{0}^{\bf K}(\epsilon) \int_{-\infty}^{+\infty} d\omega A_{\bf K}^{2}(\epsilon,\omega)(\frac{-df}{d\omega})
\ee
     and the dc resistivity is just $\rho_{xx}(U/t,T)=1/\sigma_{xx}(U/t,T)$.  
\begin{figure}
\includegraphics[width=1.1\columnwidth , height= 
1.1\columnwidth]{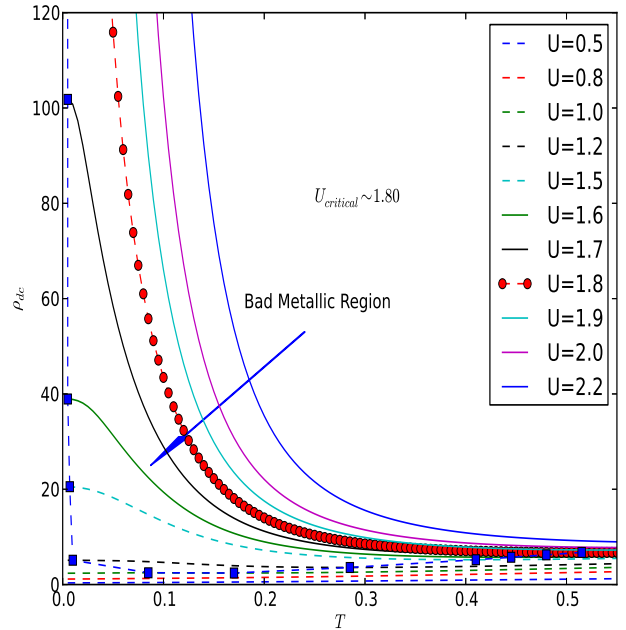} 
\caption{(Color online) The $dc$ resistivity for various $U/t$ across the continuous ``Mott'' transition in the binary-alloy disorder model.  When $0.90\leq U/t< 1.8$, an intermediate ``bad insulator'' regime separates the high-$T$ incoherent metal from the $T\rightarrow 0$ very bad metal, beyond which a split-band (``Mott'') insulator obtains.}
\label{fig:fig1}
\end{figure}
   In Fig.\ref{fig:fig1}, we show $\rho_{xx}(U/t,T)$ as $U/t$ is raised from small to large values across a critical value, $(U/t)_{c}=1.80$, where a {\it continuous} MIT occurs in the FKM within CDMFT~\cite{ourfirstpaper}.  Several features clearly stand out: $(1)$ at high $T$, $\rho_{xx}(T)~AT$ with small $A$, and always attains bad-metallic limiting values $\forall$ $U/t\geq 0.5$.  This behavior persistis up to rather low $T~0.01-0.02t$, below which it levels off to a $T$-independent value, as expected of a weakly disordered metal.  Thus, the metallic state is {\it never} a strict Landau Fermi liquid $(2)$ Remarkably, $\forall$
$U/t\geq 0.90$, $\rho_{xx}(T)$ develops a minimum at intermediate-to-low $T$, and further, $\rho_{xx}(T\rightarrow 0) > (2-3)\hbar/e^{2}$, exceeding the Mott-Ioffe-Regel (MIR) limit.  This describes a {\it re-entrant} ``transition'' from ``bad insulator'' to bad-metal at very low $T$.  Both $\rho_{xx}(T) \simeq T$ and bad-metallicity are found for the FKM in DMFT~\cite{freericks}, though we find much cleaner linear-in-$T$ behavior up to much lower $T$ here $(3)$ Even more surprisingly, in the regime $0.90\leq U/t \leq 1.80$, $\rho_{xx}(T)$ crosses over smoothly from a high-$T$ bad-metallic behavior to a progressively wider intermediate-to-low $T$ window where it shows progressively insulating behavior, followed by a second ``re-entrant transition'' to an extremely bad metal with $\rho_{xx}(T\rightarrow 0) \simeq O(20-250)\hbar/e^{2}$,
before the $T\rightarrow 0$ ``Mott'' insulating state obtains as a divergent resistivity.  These features are very different from expectations based on WL approaches, and cry out for deeper understanding.

   Theoretically, two-site CDMFT reliably captures arbitrarily strong, repeated scattering processes off spatially separated scatterers on the cluster length scale $l\simeq k_{F}^{-1}$.  Thus, it works best in the MIR regime, where $k_{F}l\simeq O(1)$, opposite to the weak-scattering regime, where $k_{F}l\gg1$.  Hence, quantum criticality in this regime has no reason tobe of the WL type, since no $(1/k_{F}l)$-expansion is now tenable.  Rather, as in the locator expansion~\cite{sudip1997}, one expects criticality associated with ``strong localization''.  To unearth the nature and effects of underlying quantum criticality, we analyze our results by performing a detailed scaling 
analysis, which we now describe. 
\begin{figure}
\includegraphics[width=1.1\columnwidth , height= 
1.1\columnwidth]{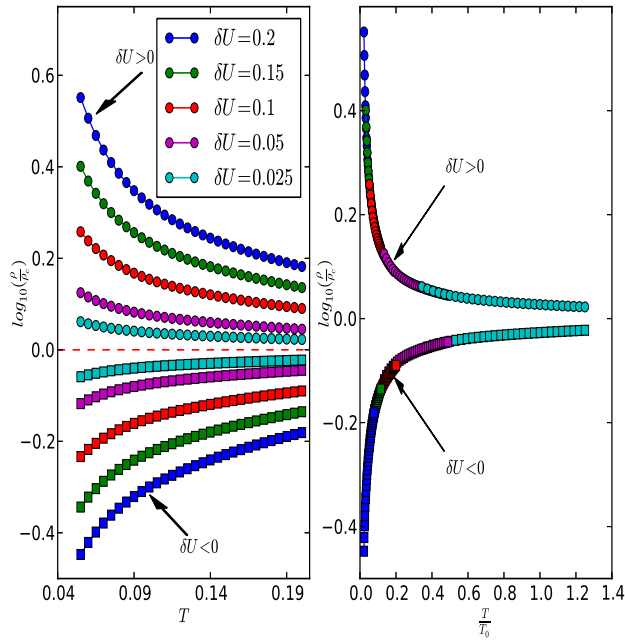} 
\caption{(Color online) Log($\rho/\rho_{c}$) vs $T$ (left panel) and Log$(\rho/\rho_{c})$ vs $T/T_{0}(U/t)$ (right panel) for same parameters as in Fig.~\ref{fig:fig1}.  Beautiful mirror symmetry around $(U/t)_{c}=1.8$ and collapse of the $T\rightarrow 0$ ``metallic'' and insulating curves on to two universal scaling trajectories is clear.}
\label{fig:fig2}
\end{figure}
 In Fig.\ref{fig:fig2}, we show log$(\rho_{xx}(T)/\rho_{xx}^{(c)}(T))$ versus $T$, where $\rho_{xx}^{c}(T)$ is the {\it critical} resistivity just at $(U/t)_{c}=1.80$ where the MIT
occurs.  Beautiful ``mirror'' symmetry of the curves about that for $(U/t)_{c}$ is testimony to the genuine quantum criticality underlying the resistivities.  Interestingly, in stark contrast to the Hubbard model (within DMFT) where $\rho_{xx}^{(c)}(T)$ is bad-metallic but quasilinear in $T$, $\rho_{xx}^{(c)}(T)|_{(U/t)_{c}}$ in the FKM is insulator-like up to very low $T$ and reaches extremely high values $O(200)\hbar/e^{2}$, attesting to very different underlying behavior.  To further unveil the novel quantum criticality, we $(1)$ plot log$(\rho_{xx}(T)/\rho_{xx}^{c}(T))$  versus the ``scaling variable'' $|U-U_{c}|/T^{1/z\nu}$ in Fig. \ref{fig:fig3} using standard procedure~\cite{kravchenko,dobrosavljevic}.  This unbiased procedure has the advantage of directly explicitly yielding $z\nu$, the product of the critical exponents associated with diverging spatial and temporal correlations at the Mott QCP, directly from the $U$-dependence of a low-energy scale, $T_{0}(U)$, which vanishes precisely at the MIT.  Remarkably, as Fig.\ref{fig:fig2}(right panel), clearly shows, we find that the ``metallic'' and insulating curves cleanly collapse on to two universal scaling curves for a wide range of $|U-U_{c}|$.  In Fig.\ref{fig:fig3}(left panel), we also confirm that $T_{0}(\delta U)\simeq c_{1}|\delta U|^{z\nu}$ with $z\nu = 1.3$.  Further, by plotting the dc conductivity as a function of $U$ in Fig.\ref{fig:fig3}(right panel), we also find that $\sigma_{xx}(U)\simeq |U_{c}-U|^{1.3}$ as the MIT is approached from the metallic side. 
\begin{figure}
\includegraphics[width=1.1\columnwidth , height= 
1.1\columnwidth]{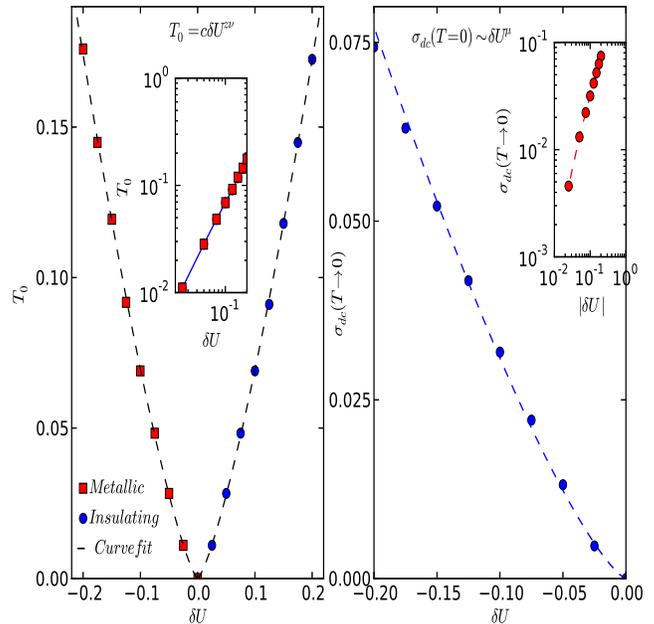} 
\caption{(Color online) The parameter $T_{0}(U/t)$ vs $\delta U=(|U-U_{c}|$ (left panel) and conductivity $\sigma_{xx}$ vs $U$.  Insets show that $T_{0}(\delta U)=(\delta U)^{1.32}\simeq (\delta U)^{4/3}$ and $\sigma_{xx}=(U_{c}-U)^{1.31}\simeq (U_{c}-U)^{4/3}$, testifying to clear quantum critical behavior (see text).}
\label{fig:fig3}
\end{figure}
 Since the critical behavior of the dc conductivity at the MIT reflects the critical divergence of the only relevant length scale, namely the localization length, $\xi(U)$, via 
$\sigma_{xx}\simeq e^{2}/\hbar\xi$~\cite{ravin}, and since $\xi(U)\simeq |U-U_{c}|^{-\nu}$, we directly extract $\nu=1.3\simeq 4/3$ and $z=1$.  It is interesting to note that $\nu=4/3$ is characteristic of a percolation mechanism for transport.  This intriguing possibility indeed holds qualitatively in the FKM as follows: as shown by Pastor {\it et al.}~\cite{pastor}, one can define a configuration averaged charge-glass susceptibility, $\chi^{[2]}$, which is also singular in the disordered ``Mott'' insulating phase of the FKM.  Noticing that inter-site correlations already effectively 
arise in our 
two-site CDMFT (near the transition on the metallic side, these read $H_{res}\simeq J_{1}\sum_{\langle i,j\rangle}\sigma_{i}^{z}\sigma_{j}^{z} +$ 4th-order Ising ``ring'' exchange for the FKM, with $\sigma_{i}^{z}=(n_{i,c}-n_{i,d})/2$ and $J_{1}\simeq 4t^{2}/U$) at sizable $U\leq U_{c}$, one expects an {\it effective} inter-site term, $H'\simeq j\sum_{\langle i,j\rangle}n_{i,c}n_{j,c}$ with a modified $j\neq J$ to persist somewhat in to the very bad metallic regime.   Since the glass transition is also signaled by the equation $(1-j\chi^{[2]})=0$, $\chi^{[2]}$will already diverge before the MIT. Thus, our finding of $\nu=4/3$ maybe due to onset of an {\it electronic} glassy dynamics near the MIT. Percolative transport is a strong possibility in glassy systems.  Though our results suggest such an emerging scenario near the MIT, clinching this link requires deeper analysis alike that by Pastor {\it et al.}, which we leave for future work.  Moreover, noticing that the Harris criterion, $\nu>2/D$, always holds for $D \geq 2$ in our case also implies that {\it intrinsic} disorder effects in the FKM cannot lead to droplet formation (which requires $\nu < 2/D$~\cite{belitz} for a second-order transition)~\cite{belitz}.  Thus, the quantum criticality is ``clean''.
\begin{figure}
\includegraphics[width=1.1\columnwidth , height= 
1.1\columnwidth]{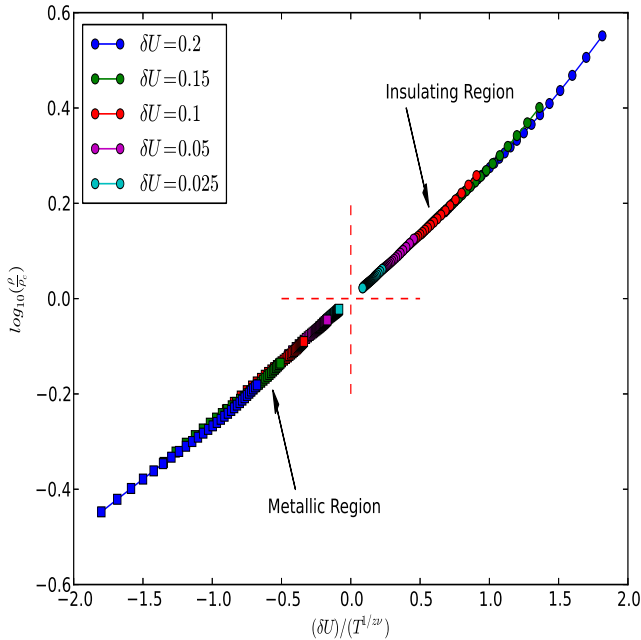} 
\caption{(Color online) Log$(\rho/\rho_{c})$ vs the scaling parameter $(\delta U)/T^{1/z\nu}$ on both sides of the MIT.  Both metallic and insulating branches exhibit the same scaling form on either side of $U_{c}$.
Continuity of the scaling curve across $U_{c}$ clearly bares "Mott" quantum criticality.}
\label{fig:fig4}
\end{figure}
Interestingly, along with the extended "mirror" symmetry, 
our $z\nu\simeq 1.3$ is qualitatively consistent with $z\nu\simeq 1.6$ for the $2D$-electron gas (2DEG) in Si near the MIT~\cite{kravchenko} and $z\nu \simeq 1.5$ for Bi films.  Our computed $z\nu=1.3$ is vey different from $z\nu=0.67$ for the one-band Hubbard model within DMFT.  The latter value is consistent with data for $2D$-organics~\cite{terletska}.  Thus, one may conclude that MITs in the 2DEG in Si and Bi films, among others, are better understood by a ``strong localization'' limit in a physical picture where strong disorder is more relevant than local Hubbard correlations.   

\begin{figure}
\includegraphics[width=1.1\columnwidth , height= 
1.1\columnwidth]{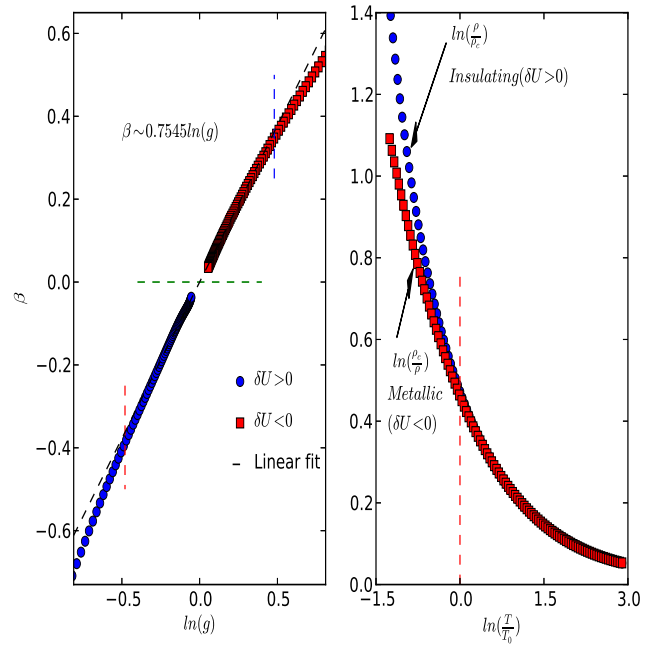} 
\caption{(Color online) The conductivity $\beta$ function vs log$(g)$ (left panel) and ln$(\rho/\rho_{c})$ (insulating) and ln$(\rho_{c}/\rho)$ (metallic) vs ln$(T/T_{0})$ (right panel).  Left panel shows that 
$\beta (g)=log(g)$ clearly holds over an extended regime in $U/t$ on both sides of $U_{c}$, testifying to clear ``Mott'' quantum criticality.  Right panel establishes the symmetry relation ln$(\rho(\delta U)/\rho_{c})=$ln$(\sigma_{xx}(-\delta U)/\sigma_{c})$ around $U_{c}$.  }
\label{fig:fig5}
\end{figure}

   Further, upon plotting the transport ${\it beta}$-function (or Gell-Mann Low function), defined by $\beta(g)=\frac{d[log(g)]}{d[log(L)]}=\frac{d[log(g)]}{d[log(T)]}$ (since $L\simeq T^{-z}$ with $z=1$ as 
above) versus log$(g)$ in Fig.~\ref{fig:fig5}(left panel), we find that $\beta(g)\simeq log(g)$ over a wide range of $U$, from the insulator, through $U_{c}$, extending deep into the ``metallic'' phase. 
 In fact, it persists up to 
$(U/t)\simeq 0.90$, showing now that the intermediate-to-low-$T$ pseudogap feature in $\rho_{xx}(T)$ in Fig.~\ref{fig:fig1} is a manifestation of this underlying Mott quantum criticality.  It is clear that this 
scaling is natural deep in the insulator, where $\rho_{xx}(T)\simeq $exp$(E_{g}/k_{B}T)$.  Its persistence deep into the metallic regime shows that the appearance of the very low-$T$ ``re-entrant metal'' 
is due to the same physical processes which cause Mott insulating behavior, and provides deeper insight into the origin of this anomalous state.  Specifically, this means that this QCP arises from 
strong-coupling physics, and is out of scope of perturbative weak-coupling schemes, as alluded to earlier.  This has additional deep implications as follows.  $(i)$ Consequent to the above, we find that the ``symmetry''
relation linking $\rho$ and $\sigma_{xx}$ on two sides of the MIT, $\frac{\rho(\delta U)}{\rho_{c}}=\frac{\sigma_{xx}(-\delta U)}{\sigma_{xx}^{c}}$, also holds over an extended region around $U_{c}$, as shown in the right panel of Fig.~\ref{fig:fig5}.$(ii)$ We also find that $log(\rho/\rho_c)$ is a universal function of the "scaling parameter" $\frac{\delta U}{T^{1/z\nu}}$.$(iii)$ Further, this also allows us to explicitly construct $\beta(g)$ for a specific microscopic model (known to be a hard task)~\cite{sudip1997} as follows: In scaling approaches to 
WL~\cite{LeeTVR1985}, 
$\beta(g)$ depends explicitly (only) on $g$, and that the probability distribution of $g$, $P(g)$, is sharply peaked at its mean value.  This assumption breaks down at ``strong'' localization, where one 
expects a broad distribution, $i.e$, $P(g)$ is broad.  It has been argued, based on insight from a locator expansion~\cite{sudip1997}, that it is $P($log$g)$, or more generally $P[$log$\phi(g)]$ 
with $\phi(g)=a/g+b+cg+...$ as $g\rightarrow 0$ that is sharply peaked in this case.  Then it turns out that $\beta(g)\simeq$ log$[\phi(g)/\phi(g_{c})]$, with $g_{c}$ the critical conmductivity.  Comparing 
this with our results, we now explicitly find that $\phi(g)\simeq 1/g$ for the FKM.  Finally, it is interesting that similar scaling features are also seen in DMFT (see Supplementary Information ({\bf SI}), but with $z\nu=1.2$, distinct from $z\nu=4/3$ found in CDMFT.  Thus, all conclusions found above remain valid, and the only important difference is that the glassy dynamics strongly hinted at in CDMFT (see above) is absent in DMFT.
         
   Thus, our findings provide clinching support for clear manifestations of an unusual quantum criticality associated with the continuous Mott-like MIT.  Perfect mirror symmetry, along with 
$\beta(g)\simeq -ln(g)$ and its persistence deep into the ``metallic'' regime all indicate similarities with Mott criticality in the Hubbard model.  But while such features appear above 
the finite-$T$ end-point ($T^{*}$) of the line of first-order Mott transitions in the HM, they persist down to $T=0$ in the FKM, underlining a genuine ``Mott'' QCP.  We can understand this qualitatively as 
follows: observe that the Landau quasiparticle picture  is already destroyed above $T_{LFL}< T^{*}$ in the HM~\cite{terletska}.  We are then left with a bad-metal where absence of coherent $\downarrow$-spin 
recoil in the HM prevents the lattice Kondo effect, making it possible to ``map'' the HM onto two coupled FKMs (one for each spin species)~\cite{edwards}.  This qualitatively explains why the ``Mott'' 
criticality features we find for the FKM resemble those seen for the HM, even though $(z\nu)^{FKM}\simeq 1.3\simeq 2(z\nu)^{HM}\simeq 0.67$.  We are presently unable to explain this difference.  
Experimentally, we posit that this QCP leaves leaves its imprint in $\rho_{xx}(T)$ as a ``bad insulator'' at intermediate $T$, followed by an anomalously bad metal as $T\rightarrow 0$: this is also distinct 
from the Hubbard case, where a bad-metallic $\rho_{xx}^{c}(T)\simeq AT$ obtains at the critical point.  Finally, within (C)DMFT, the quantum disordered phase in the FKM is known to possess a finite residual 
entropy $O($ln$2)$ per site.  Along with infra-red branch-cut continuum spectral functions~\cite{ourfirstpaper} in earlier work, our findings are reminiscent of ``hologhraphic duality'' scenarios~\cite{subir}.
  Thus such novel quantum criticality, originally proposed for QCPs associated with Kondo-destruction approaches to ($T=0$) melting of quasiclassical order, may also hold for ``Mott'' quantum criticality 
associated with a {\it continuous} metal-insulator transition.  

%\vspace{0.5cm}
\acknowledgements

\newpage
{\bf Supplementary Information(single site dmft):}

   In this section, we exhibit a remarkable feature.  Most, but not all, novel features found in our two-site CDMFT study of transport are already visible in single-site DMFT.  In particular, at first glance, Figs $6,7,8$ and $9$ cleanly exhibit all features seen in two-site CDMFT calculations. 
\begin{figure}[h]
\includegraphics[width=1.0\columnwidth , height= 
1.0\columnwidth]{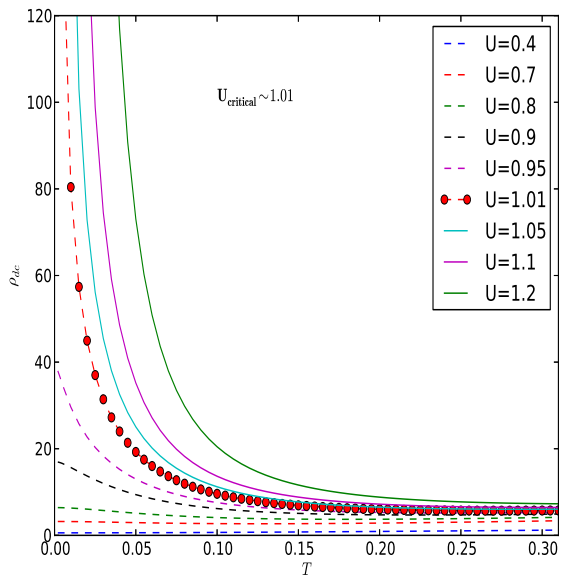} 
\caption{(Color online) $dc$ resistivity as a function of $U/t$ in single-site DMFT.  Though the trend is similar to that in CDMFT, $\rho_{dc}(U_{c},T)$ is $O(40\hbar/e^{2})$, much smaller than $O(250\hbar/e^{2})$ seen in the CDMFT.}
\label{fig:fig6}
\end{figure}

   To obtain these features, we simply use the exact DMFT spectral function, $A({\bf k},\omega)=-1/\pi$Im$G({\bf k},\omega)$ with $G({\bf k},\omega)^{-1}=\omega-\epsilon_{\bf k}-\Sigma(\omega)$.  The exact DMFT self-energy reads

\be
\Sigma(\omega)=U\langle n_{i,d}\rangle + \frac{U^{2}\langle n_{i,d}\rangle(1-\langle n_{i,d}\rangle)}{\omega-U(1-\langle n_{i,d}\rangle) -t^{2}G_{loc}(\omega)}
\ee 
for the Bethe lattice.  We insert the $A({\bf k},\omega)$ 
 in the Kubo formula for the current-current correlation function.  It is well known that this is literally exact, since irreducible vertex corrections in the Bethe-Salpeter equations for conductivities rigorously drop out in this limit.
\begin{figure}[ht]
\includegraphics[width=1.0\columnwidth , height= 
1.0\columnwidth]{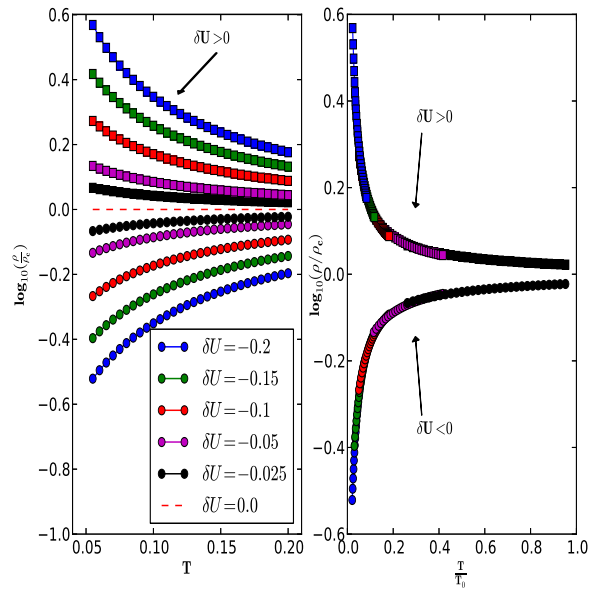} 
\caption{(Color online) Log$(\rho/\rho_{c})$ versus $T$ (left panel) and versus $T/T_{0}$ (right panel) exhibiting clear  mirror symmetry about $U_{c}^{DMFT}/t=1.01$.  This is very similar to CDMFT results, indicating that no qualitative changes occur for the scaling features upon use of DMFT.}
\label{fig:fig7}
\end{figure} 
   Upon closer inspection, however, some important distinctions between DMFT and CDMFT results arise.

   $(i)$ The $dc$ resistivity follows a similar trend with $U/t$.  However, it attains values $O(40)\hbar/e^{2}$ just before the MIT, in contrast to the much larger values $O(250)\hbar/e^{2}$ in CDMFT (see main text).

\begin{figure}[h]
\includegraphics[width=1.\columnwidth , height= 
1.\columnwidth]{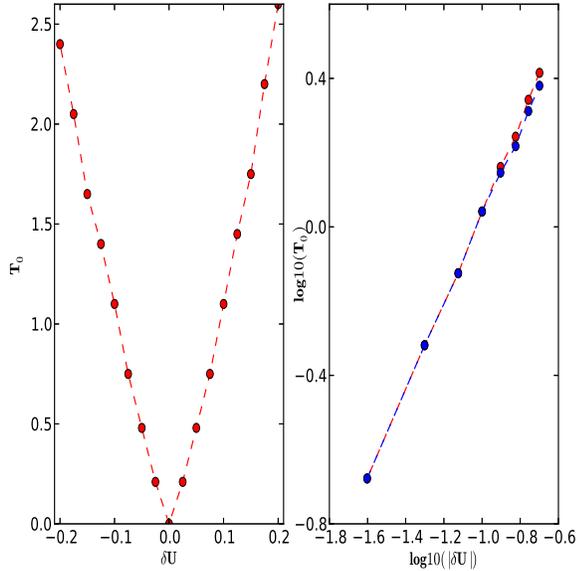} 
\caption{(Color online) $T_{0}(\delta U)$ as a function of $\delta U$ on normal (left panel) and on log-log scale (right panel).  Though behavior qualitatively very similar to CDMFT results obtains, the exponent product
$z\nu=1.2$ (right panel) in contrast to $z\nu=4/3$ in CDMFT.  This means that onset of glassy dynamics does not
get reflected in single-site theories.}
\label{fig:fig8}
\end{figure}

   $(ii)$  Though similar scaling of $T_{0}(\delta U)$ and $\beta(g)$ of comparable good quality is seen in DMFT results as well, the DMFT finding of $z\nu=1.2$ is different from $z\nu=4/3$ found in CDMFT.  The latter is the expected value for a classical percolation regime associated with onset of electronic {\it glassy} dynamics near the MIT.  Thus, our results point toward the need for a cluster extension of DMFT to access onset  of a dynamical glassy regime in transport close to the MIT.

\begin{figure}
\includegraphics[width=1.0\columnwidth , height= 
1.\columnwidth]{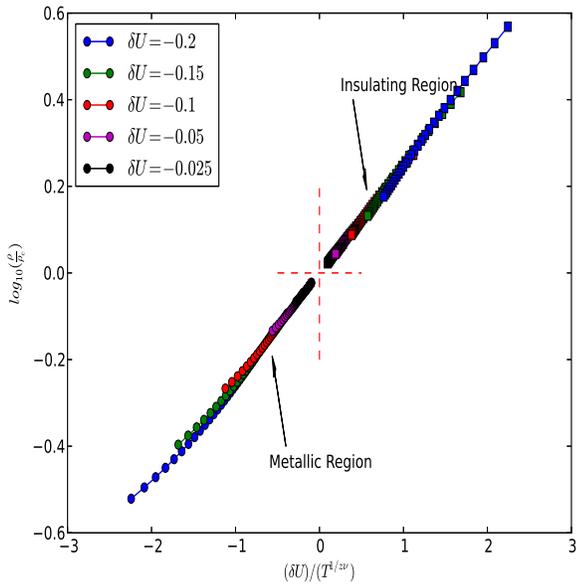} 
\caption{(Color online)  Log$(\rho/\rho_{c})$ versus the scaling parameter $\frac{\delta U}{T^{1/z\nu}}$ for the metallic and insulating phases in DMFT.  Continuity across the MIT and ideentical form of the scaling function on both sides of the MIT testify to Mott quantum criticality in the FKM within DMFT as well.}
\label{fig:fig9}
\end{figure}

   $(iii)$ finally, since $(z\nu)_{DMFT}=1.2 > 2/D$ for $D\geq 2$ satisfies the Harris criterion as well, 
no droplet formation can occur, and the quantum criticality we find is ``clean''.

%\begin{figure}
%\includegraphics[width=1.1\columnwidth , height= 
%1.1\columnwidth]{fig10.eps} 
%\caption{(Color online) }
%\label{fig:fig10}
%\end{figure}


\begin{thebibliography}{18}

\bibitem{pwa1958pwa1979}  P. W. Anderson, Phys. Rev. {\bf 109}, 1492 (1958); E. Abrahams {\it et al.}, Phys. Rev. Lett. {\bf 42}, 673 (1979).

\bibitem{LeeTVR1985} P. Lee and T. V. Ramakrishnan, Revs. Mod. Phys.  {\bf 57}, 287 (1985).

\bibitem{kravchenko} S. Kravchenko {\it et al.}, Phys. Rev. B {\bf 51}, 7038 (1995)

\bibitem{arindam} M. Pepper {\it et al.}, Physical Review Letters {\bf 100}, 016805 (2008).

\bibitem{rosenbaum} S. Field and T. Rosenbaum, Phys. Rev. Lett. {\bf 55}, 522 (1985).

\bibitem{dobrosavljevic}  A. Punnoose and A.M. Finkel'stein, Science, {\bf 310}, 289 (2005); 
V. Dobrosavljevic, arXiv:1602.00131 (Contribution to: "Strong Correlation Phenomena around 2D Conductor-Insulator Transitions", edited by S. V. Kravchenko, Pan Stanford Publishing, 2016) and references therein.

\bibitem{terletska} H. Terletska {\it et al.}, Phys. Rev. Lett. {\bf 107}, 026401 (2011) 

\bibitem{sudip1997} V. Dobrosavljević {\it et al.}, Phys. Rev. Lett. {\bf 79}, 455 (1997). 

\bibitem{freericks} J. K. Freericks and V. Zlatić, Rev. Mod. Phys. {\bf 75}, 1333 (2003). 

\bibitem{ourfirstpaper} P. Haldar, M. S. Laad and S. R. Hassan, arXiv:1603.00301.

\bibitem{kotliar} K. Haule and G. Kotliar, Europhys. Lett. {\bf 77}, 27007 (2007).

\bibitem{rowlands} D. A. Rowlands, J. B. Staunton, and B. L. Györffy, Phys. Rev. B {\bf 67}, 115109 (2003).

\bibitem{jarrell} M. Jarrell and H. R. Krishnamurthy, Phys. Rev. B {\bf 63}, 125102 (2001).

\bibitem{ravin} S. Bogdanovich, M. P. Sarachik, and R. N. Bhatt, Phys. Rev. Lett. {\bf 82}, 137 (1999).
Phys. Rev. Lett. 82, 137

\bibitem{pastor} V. Dobrosavljevi ́c, D. Tanaskovi ́c, and A. A. Pastor, Phys. Rev. Lett. {\bf 90}, 016402(4), (2003).

\bibitem{belitz} T.R. Kirkpatrick, D. Belitz, arXiv:1602.01447.

\bibitem{edwards} D. M. Edwards, J. Phys. condens. Matter. {\bf 5}, 161 (1993). 

\bibitem{subir} S. Sachdev, Phys. Rev. Lett. {\bf 105}, 151602 (2010). 
\end{thebibliography}
\end{document}